\documentclass{CUP-JNL-DTM}%

\usepackage{graphicx}
\usepackage{multicol,multirow}
\usepackage{amsmath,amssymb,amsfonts}
\usepackage{mathrsfs}
\usepackage{amsthm}
\usepackage{rotating}
\usepackage{appendix}
\usepackage[numbers]{natbib}
\usepackage{ifpdf}
\usepackage[T1]{fontenc}
\usepackage{newtxtext}
\usepackage{newtxmath}
\usepackage{textcomp}
\usepackage{xcolor}
\usepackage{lipsum}
\usepackage[colorlinks,allcolors=blue]{hyperref}

\theoremstyle{definition}

\numberwithin{equation}{section}

\jname{Environmental Data Science}
\articletype{Application paper}
\jyear{2024}

\begin{document}

\begin{Frontmatter}

\title[Article Title]{Simulating the Air Quality Impact of Prescribed Fires Using Graph Neural Network-Based PM$_{2.5}$ Forecasts}

\author[1]{Kyleen Liao}
\author[2]{Jatan Buch}
\author[2]{Kara Lamb}
\author[2]{Pierre Gentine}

\authormark{Liao \textit{et al}.}

\address[1]{\orgdiv{Saratoga High School}, \orgaddress{\city{Saratoga}, \postcode{95070}, \state{CA},  \country{USA}}, \email{kyleenliao@gmail.com}}

\address[2]{\orgdiv{Department of Earth and Environmental Engineering, Columbia University},  \orgaddress{\city{New York City}, \postcode{10027}, \state{NY},  \country{USA}}}



\abstract{The increasing size and severity of wildfires across the western United States have generated dangerous levels of PM$_{2.5}$ concentrations in recent years. In a changing climate, expanding the use of prescribed fires is widely considered to be the most robust fire mitigation strategy. However, reliably forecasting the potential air quality impact from prescribed fires, which is critical in planning the prescribed fires' location and time, at hourly to daily time scales remains a challenging problem. In this paper, we introduce a spatial-temporal graph neural network (GNN) based forecasting model for hourly PM$_{2.5}$ predictions across California. Using a two-step approach, we leverage our forecasting model to estimate the PM$_{2.5}$ contribution of wildfires. Integrating the GNN-based PM$_{2.5}$ forecasting model with prescribed fire simulations, we propose a novel framework to forecast the PM$_{2.5}$ pollution of prescribed fires. This framework helps determine March as the optimal month for implementing prescribed fires in California and quantifies the potential air quality trade-offs involved in conducting more prescribed fires outside the fire season.
}

\end{Frontmatter}

\section*{Impact Statement}
PM$_{2.5}$ pollution poses significant health risks and is responsible for millions of deaths per year. Our work forecasts the PM$_{2.5}$ concentration at sparse sensor locations and estimates the fire-specific PM$_{2.5}$ contribution to assess their impact on air quality. Furthermore, prescribed fires, while preventing wildfires also generate PM$_{2.5}$, raising concerns about the air quality trade-offs. To the best of our knowledge, our work is the first to apply machine learning to predict the PM$_{2.5}$ concentration from simulated prescribed fires. We use our forecasting model to conduct novel experiments that can help the fire service better understand and minimize the pollution exposure from prescribed fires. 


\section[Introduction]{Introduction}

Across many parts of the western United States (WUS), wildfire size, severity, and fire season length have increased due to climate change \cite{williams2019}. Wildfires across the WUS have led to the largest daily mean PM$_{2.5}$ (particulate matter < 2.5 microns) concentrations observed by ground-based sensors in recent years \cite{burke2021}, and exposure to PM$_{2.5}$ is responsible for 4.2 million premature deaths worldwide per year \cite{who}. Within California, additional PM$_{2.5}$ emissions from extreme wildfires over the past 8 years have reversed nearly two decades of decline in ambient PM$_{2.5}$ concentrations \cite{burke2023}. 

Due to the numerous severe health consequences from PM$_{2.5}$ pollution exposure, performing accurate and temporally fine-grained PM$_{2.5}$ predictions has become increasingly significant. A recent study by Aguilera et al. (2021) \cite{aguilera2021} found the PM$_{2.5}$ emitted from wildfires to be more toxic than the PM$_{2.5}$ emitted from ambient sources. Accurate PM$_{2.5}$ predictions are also important in the context of prescribed fires, or controlled burns, which have been widely accepted as an effective land management tool and could have the potential to reduce the resulting smoke from future wildfires \cite{kelp2023}. Since air quality is a major public concern surrounding prescribed fires \cite{mccaffrey2006}, land managers conducting these burns require access to robust, near real-time predictions of downwind air pollution in order to determine suitable locations and burn windows. 

However, most data-driven PM$_{2.5}$ forecasting algorithms do not distinguish between the ambient PM$_{2.5}$ concentration and the additional concentration due to fire emissions. Additionally, previous work studying the effect of prescribed fires on pollution used chemical transport models (CTMs) like the Community Multiscale Air Quality (CMAQ) and Goddard Earth Observing System Atmospheric Chemistry (GEOS-Chem) models to calculate the PM$_{2.5}$ impact of prescribed fires at different locations \cite{kelp2023}. While CTMs can model the chemical processes in PM$_{2.5}$ transport, generating accurate predictions is computationally intensive because of the complex chemical interactions. Furthermore, the extensive calculations in CTMs make it challenging to explore a large range of parameters for simulating prescribed burns \cite{askariyeh2020, byun2006, zaini2022}. In air quality predictions, machine learning models have been shown to outperform CTMs in terms of accuracy and computational burden \cite{rybarczyk2018}. While several studies have used machine learning to forecast air quality \cite{wang2020PM25GNN, li2023}, this is the first research paper, to the best of our knowledge, that utilizes machine learning to predict the PM$_{2.5}$ concentration from simulated prescribed fires.

Our research builds upon the graph neural network (GNN) machine learning model from Wang et al. (2020) \cite{wang2020PM25GNN}, which was used to forecast non-wildfire-influenced PM$_{2.5}$ pollution in China. In contrast, our work focuses on predicting fire-influenced PM$_{2.5}$ in California. The spatio-temporal modeling capabilities of GNNs coupled with domain knowledge make the model especially valuable for PM$_{2.5}$ prediction with spatially sparse monitor observations, enabling the GNN to outperform baseline machine learning architectures such as long short-term memory (LSTM) and multilayer perceptron (MLP) neural networks. This paper focuses on predicting PM$_{2.5}$ pollution at an hourly resolution in California and its two applications: 1) quantifying fire-specific PM$_{2.5}$ concentration and 2) forecasting the pollution levels emitted from simulated prescribed fire events. Specifically, we incorporated satellite-derived data on fire intensity within a GNN model to forecast the PM$_{2.5}$ concentration from ambient sources, observed fires, and simulated controlled burns. Our GNN-based forecasting framework can help policymakers better isolate the PM$_{2.5}$ concentration emitted from wildfires. Additionally, our forecasts can aid land managers in minimizing the PM$_{2.5}$ exposure of vulnerable populations during controlled burns and facilitate community discussions of potential locations and burn windows for prescribed fires.

\section{Dataset} \label{dataset}
Our dataset consists of PM$_{2.5}$, meteorological, and fire data at an hourly resolution over 5 years (2017-2021). The PM$_{2.5}$ concentration data, at a total of 112 sparse air quality sensor locations in California, is collected from both the California Air Resources Board as well as the Environmental Protection Agency \cite{carb, epa}. The MissForrest algorithm \cite{stekhoven2011} was used to impute the missing PM$_{2.5}$ observations from offline sensors. The data for the 7 meteorological variables, which include u and v horizontal components of wind, total precipitation, and air temperature, are retrieved from the ERA5 Reanalysis database \cite{hersbach2020}. The full list of predictors is in Table \ref{predictors}. Though the meteorological variables may capture the diurnal PM$_{2.5}$ cycles and seasonal patterns, the Julian date and hour of the day are also included as predictors to provide the model with additional context. 

\begin{table}[!ht] 
  \caption{GNN predictors}
  \label{predictors}
  \centering
  \begin{tabular}{lll}
    \cmidrule(r){1-3}
    Predictor Name     & Unit     & Source \\
    \midrule
    Planetary Boundary Layer Height (PBLH) & m  & ERA5 Reanalysis     \\
    u-component of wind     & m/s & ERA5 Reanalysis      \\
    v-component of wind     & m/s       & ERA5 Reanalysis  \\
    2m Temperature     & K      & ERA5 Reanalysis  \\
    Dewpoint temperature     & K      & ERA5 Reanalysis  \\
    Surface pressure     & Pa       & ERA5 Reanalysis  \\
    Total precipitation     & m       & ERA5 Reanalysis  \\
    WIDW FRP within 25km, 50km, 100km, 500km     & MW      & VIIRS  \\
    Number of fires within 500km     & 1       & VIIRS  \\
    Julian date     & 1       & N/A  \\
    Time of day     & 1       & N/A  \\
    \bottomrule
  \end{tabular}
\end{table}

\begin{figure}[!ht]
  \centering
  \includegraphics[width=0.37\textwidth]{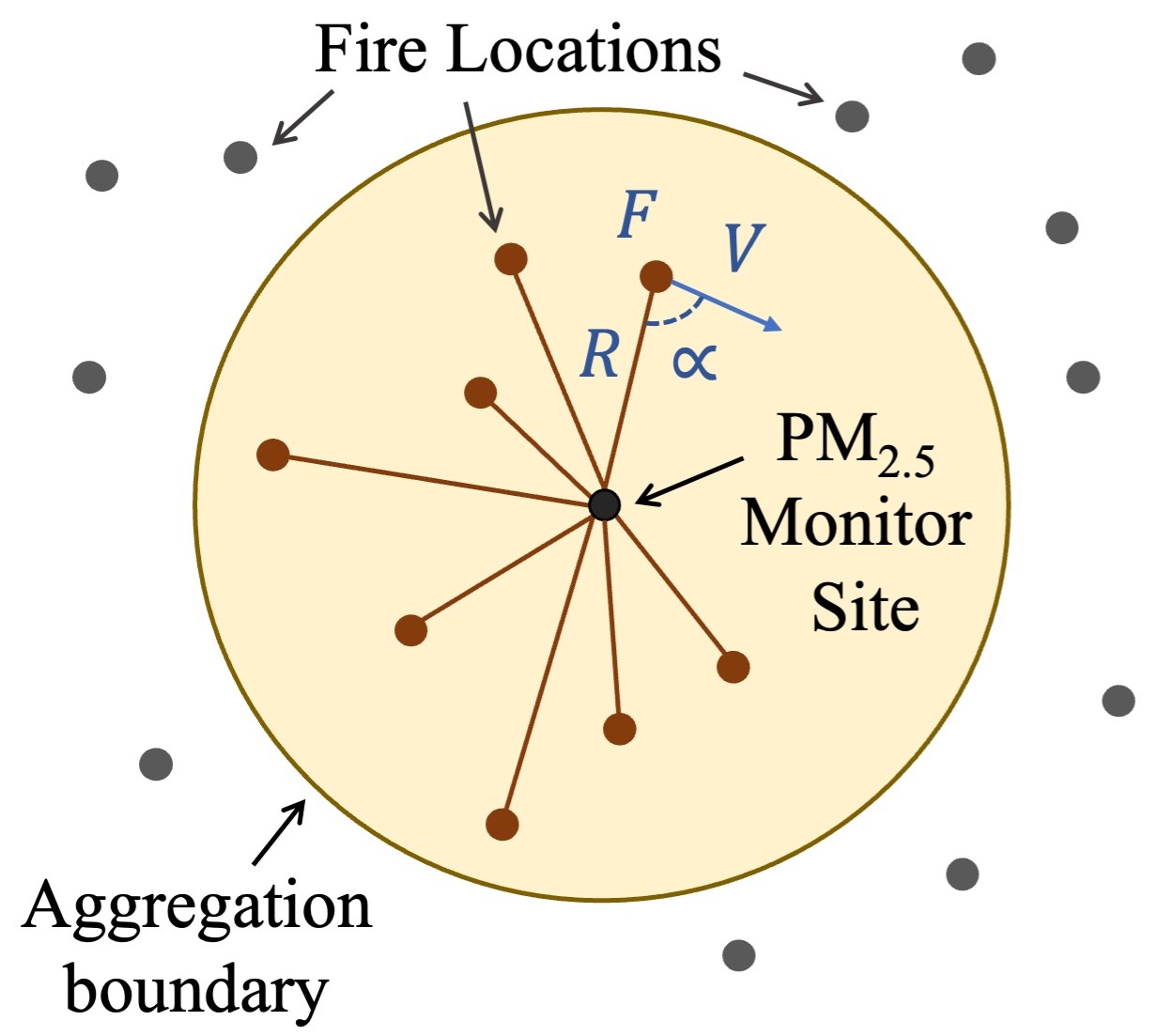}
  \caption{FRP aggregated around a given radius for each PM$_{2.5}$ monitor location using wind and distance information}
  \label{fig:aggregation}
\end{figure}

The fire radiative power (FRP) provides information about the fire intensity. The FRP at each fire location is taken from the Visible Infrared Imaging Radiometer Suite (VIIRS) \cite{schroeder2014} instrument on board the Suomi satellites. In order to assess the impact of nearby fires at the location of a PM$_{2.5}$ monitor, we aggregate the FRP values of all active fires within radii of 25km, 50km, 100km, and 500km. To emphasize the fires that would likely have a more substantial downwind effect on PM$_{2.5}$ concentration, we use inverse distance weighting (IDW) and wind-based weighting in the FRP aggregation. For each PM$_{2.5}$ monitor location, aggregations are performed on radii of 25km, 50km, 100km, and 500km to derive the wind and inverse-distance weighted (WIDW) FRP using the process described in Figure \ref{fig:aggregation} and Equation \ref{eq:1}, 

\begin{equation} \label{eq:1}
F_{\rm WIDW}= \sum_{i=1}^n \frac{F_i\left|V_i\right| \cos \left(\left|\alpha_i\right|\right)}{4 \pi R_i{ }^2}
\end{equation}

where $n$ is the number of fire locations within a certain radius of the PM$_{2.5}$ monitor site, $F$ is the FRP value at the fire location, $|V|$ is the magnitude of the wind speed at the fire location, $\alpha$ is the relative angle between the wind direction and the direction from the fire to the PM$_{2.5}$ monitor, and $R$ is the distance between the fire site and PM$_{2.5}$ monitor. The number of fires within 500km of a PM$_{2.5}$ site is also included in the dataset. The prescribed fire latitude, longitude, and duration data retrieved from the California Department of Forestry and Fire Protection (Cal Fire) \cite{calfire} is not represented as a variable in the training dataset, but instead used in Experiments 1 and 2 when simulating prescribed fires.

\section{PM$_{2.5}$ Forecasts} 
\subsection{Graph Neural Network (GNN)} 
\subsubsection{Methods}\label{gnnmethod}
\begin{figure*}
  \centering
   \includegraphics[width=\textwidth]{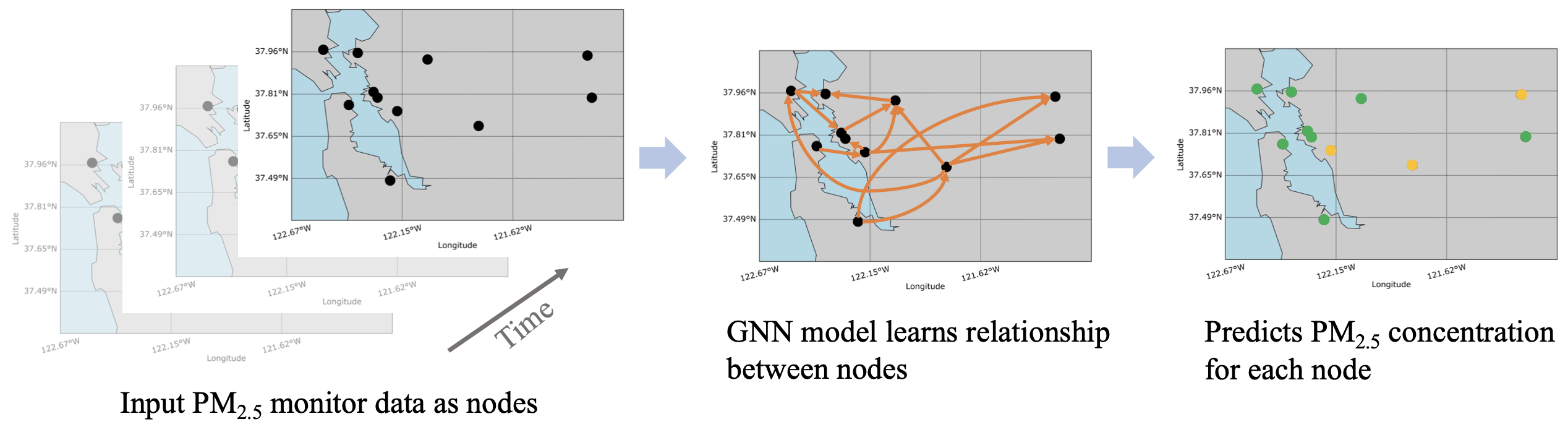}
    \caption{Graph neural network (GNN) used in our PM$_{2.5}$ forecasting model considers PM$_{2.5}$ monitors as nodes in the graph and produces node-level predictions}
    \label{fig:gnnfig}
\end{figure*}

We trained a spatio-temporal graph neural network (GNN) model from Wang et al. (2020) \cite{wang2020PM25GNN} to predict PM$_{2.5}$ concentration at an hourly temporal resolution utilizing spatially sparse observations, as illustrated in Figure \ref{fig:gnnfig}. The GNN model has a directed graph, where nodes represent the locations of PM$_{2.5}$ monitors and edges model the PM$_{2.5}$ movement and interactions between monitors, making it well suited to leverage non-homogeneous data. Thus, the node features include the meteorological and fire-related variables, and the edge attributes include the wind direction and speed at the source node and the direction and distance between any two locations. To enable the model to capture both the spatial and temporal propagation of PM$_{2.5}$, the GNN model is integrated with a recurrent neural network (RNN) component. Furthermore, domain knowledge is also incorporated in the graph representation. For instance, the graph explicitly includes wind direction information and considers geographical elevation differences. 

As shown in Table \ref{table:split}, for the GNN model, two years are used for training and one year each for validation and testing. The year 2019 is excluded during training, validation, and testing because the 2019 fire season was an outlier and was less damaging than the other years. Validating and testing the model on the years 2020 and 2021 respectively would help us gain a better understanding of the model’s performance during intense fires, as both 2020 and 2021 had severe wildfire seasons. Our model produces forecasts for a prediction window of 48 hours into the future based on a historical window of 240 hours.

\begin{table}[!h]
  \caption{Training/Validation/Testing split}
  \fontsize{9}{11}\selectfont %
  \centering
  \begin{tabular}{lll}
    \toprule
    Training     & Validation     & Testing \\
    \midrule
    1/1/2017 - 12/31/2018 & 1/1/2020 - 12/31/2020 & 1/1/2021 - 12/31/2021\\
    \bottomrule
  \end{tabular}
  \label{table:split}
\end{table}

\subsubsection{Results} \label{gnnresultsec}
To evaluate the GNN model, the GNN's performance was compared to two baseline models, the long short-term memory (LSTM) and multilayer perceptron (MLP) models. The GNN model had the lowest mean absolute error (MAE) and root mean squared error (RMSE) values, as shown in Table \ref{table:gnnresults}. As a reference for the error, very unhealthy and hazardous PM$_{2.5}$ levels are $\geq$ 150.5 $\mu$g/${\rm m}^3$. 

\begin{table}
  \fontsize{9}{11}\selectfont %
  \caption{Results of the GNN, LSTM, and MLP models}
  \centering
  \begin{tabular}{l lll}
    \toprule
    & GNN     & LSTM     & MLP\\
    \cmidrule(r){2-4}
    MAE & 5.23 & 5.73 & 6.24\\
    RMSE & 6.72 & 7.32 & 7.83\\
    \bottomrule
  \end{tabular}
\label{table:gnnresults}
\end{table}

\begin{figure}
  \centering
  \includegraphics[width=\textwidth]{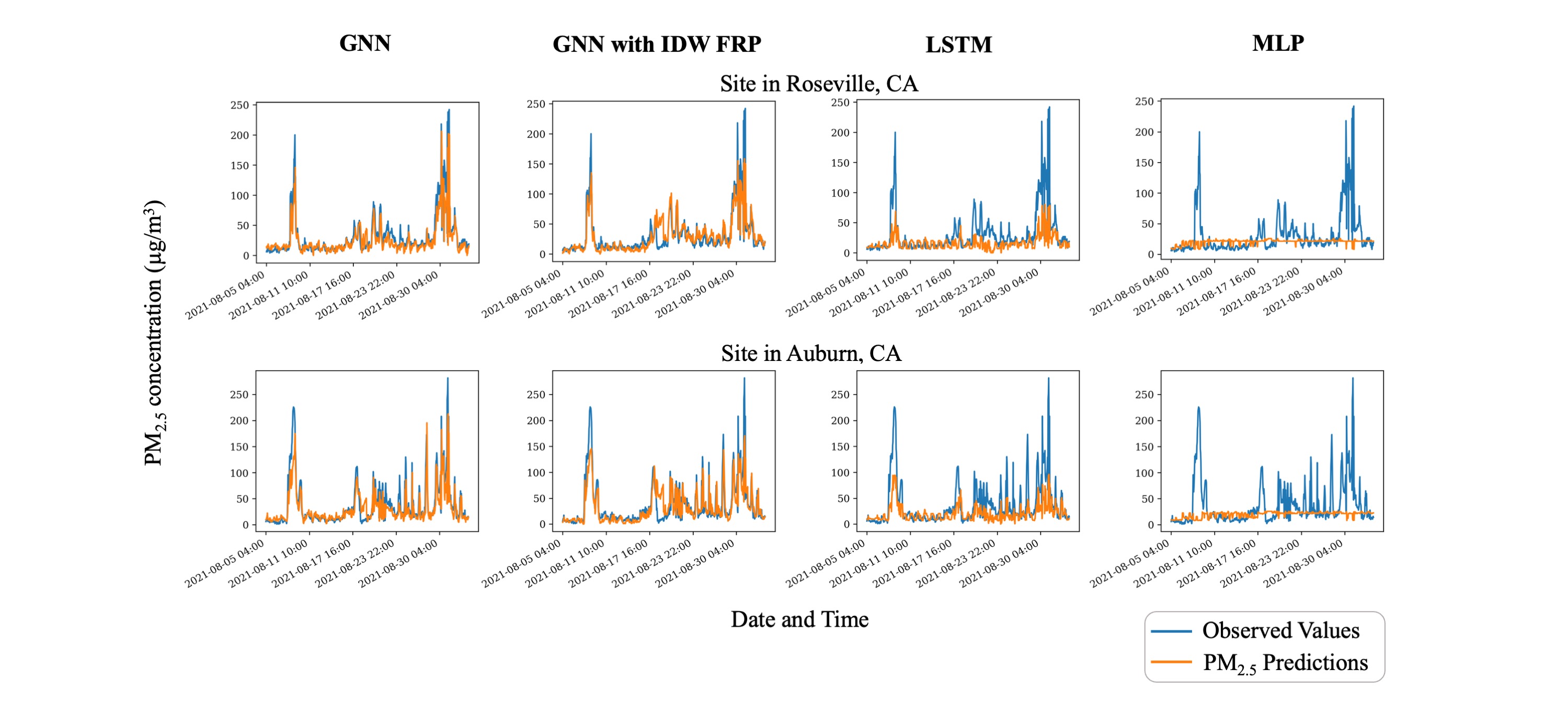}
  \caption{PM$_{2.5}$ predictions one hour into the future from a temporal subset of testing results for example sites. The GNN (column one), LSTM, and MLP all use WIDW FRP variables, while the GNN with IDW FRP (column two) uses IDW FRP variables}
  \label{fig:model_result}
\end{figure}

Additionally, the time series results of the GNN, LSTM, and MLP were graphed to analyze the results. Figure \ref{fig:model_result} displays predictions one hour into the future from the testing results of two example sites. The graphs showed that the GNN was better able to predict elevated concentrations in comparison to the MLP and LSTM. However, from the graphs of all three models, it was evident that there was a tendency for the model to output a value close to the observed value at the previous time point as its prediction. This issue is also prevalent in other studies in this area and in the broader machine-learning field. 

The GNN’s performance was also compared to the performance of a GNN trained on a dataset with only inverse-distance weighted (IDW) FRP, not wind and inverse-distance weighted (WIDW). The GNN with WIDW FRP has slightly higher MAE and RMSE, as seen in Table \ref{table:gnnwind}. However, the graphs of the results showed that the GNN with WIDW FRP was better able to predict elevated PM$_{2.5}$ values. This is significant, as current prediction models under-predict fire-influenced PM$_{2.5}$ concentration \cite{reid2021}. The reason for the slightly higher MAE and RMSE for the GNN with WIDW FRP seems to be that the model tends to slightly over-predict low-concentration values. 

\begin{table}
  \fontsize{9}{11}\selectfont %
  \caption{Results of the GNN with WIDW FRP and IDW FRP}
  \centering
  \begin{tabular}{l ll}
    \toprule
    & GNN with WIDW FRP    & GNN with IDW FRP    \\
    \cmidrule(r){2-3}
    MAE & 5.23 & 5.11 \\
    RMSE & 6.72 & 6.62 \\
    \bottomrule
  \end{tabular}
\label{table:gnnwind}
\end{table}

\subsection{Fire-specific PM$_{2.5}$ Forecasts} 
\subsubsection{Methods}\label{firespecific}

\begin{figure*}
  \centering
   \includegraphics[width=\textwidth]{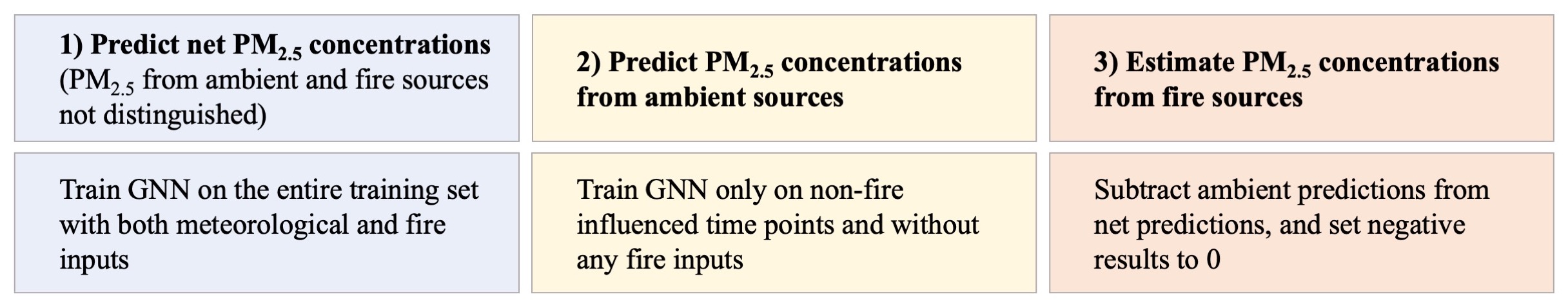}
    \caption{Conceptual diagram of the methodology for distinguishing fire-specific and ambient PM$_{2.5}$ concentrations}
    \label{fig:firespecific}
\end{figure*}

For the task of distinguishing the pollution emitted from wildfires, a two-step process is used. A GNN model is first trained to predict the total PM$_{2.5}$ concentration, and a second GNN is trained to predict the PM$_{2.5}$ emitted from ambient sources. The predictions from the ambient-focused GNN are subtracted from the forecasts from the first GNN to produce an estimate of the fire-specific PM$_{2.5}$. This process is outlined in Figure \ref{fig:firespecific}.

The detailed methodology for the first GNN model, which is trained to predict the total PM$_{2.5}$ concentration, is outlined in Section \ref{gnnmethod}. This GNN model is trained on all data variables, including meteorological and fire-related data. The second GNN model, on the other hand, focuses only on predicting the ambient PM$_{2.5}$ and is thus trained only on the meteorological data. For the second GNN, fire variables are excluded during training to prevent the model from learning the effect of fires on PM$_{2.5}$ concentration. All the data during fire events are also excluded because including time points during fires would allow the model to learn the influence of fires from meteorological variables like temperature. Therefore, for the second GNN to only predict ambient pollution, the model should only be trained on meteorological data and non-fire-influenced time points. However, selecting the time points without fire events is challenging, as PM$_{2.5}$ particles emitted by a fire can persist in the air for weeks \cite{who2006}. Performing data analysis revealed that relatively high FRP values continued to affect the PM$_{2.5}$ concentration for over a week. Thus during training, all time points within 10 days of WIDW FRP 500km values greater than 0.15 are excluded. The second GNN is then validated and tested on all time points to obtain predictions of ambient PM$_{2.5}$ for all time points, which is necessary to quantify the fire-specific PM$_{2.5}$. 

\subsubsection{Results}
\begin{figure}
  \centering
  \includegraphics[width=\textwidth]{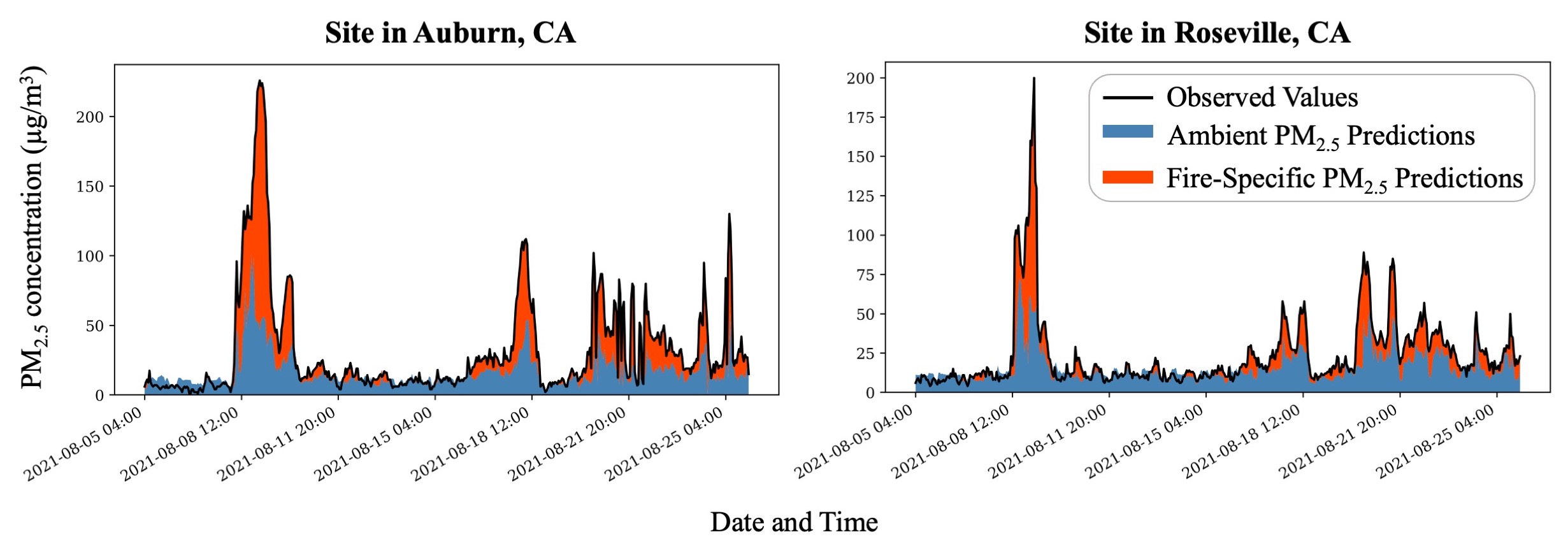}
  \caption{Ambient and fire-specific PM$_{2.5}$ predictions one hour into the future from a temporal subset of testing results for example sites}
  \label{fig:fire-specific-result}
\end{figure}

The GNN model trained to predict only ambient PM$_{2.5}$ had an MAE of 6.30 $\mu$g/${\rm m}^3$ and RMSE of 7.80 $\mu$g/${\rm m}^3$ for its predictions on time points without fire influence (times not within 10 days of WIDW FRP 500km values greater than 0.15). Figure \ref{fig:fire-specific-result} visually distinguishes the ambient and fire-specific PM$_{2.5}$ forecasts. For the fire-specific estimate, there is no ground truth value for the PM$_{2.5}$ concentration produced by ambient versus wildfire sources, and thus no metric describing the accuracy of the forecast. However, the fire-specific predictions can be assumed to have a comparable accuracy as the GNN model results for the net pollution and the ambient pollution.

\section{Prescribed Fire Simulations} \label{prescfire}
A major contribution of this work is the novel simulation of the effect of prescribed fires in conjunction with the GNN-based prediction of the resulting PM$_{2.5}$ concentrations. This framework is illustrated in Figure \ref{fig:pipeline}. The prescribed fires are simulated by transposing historical prescribed fires to target times. The Cal Fire \cite{calfire} latitude, longitude, and duration data for the prescribed burns are matched with the VIIRS FRP data. The transposed prescribed fire FRP information is combined with the observed meteorological data at the target times and inputted into the GNN model, which produces the PM$_{2.5}$ predictions.

Using this framework, we perform two model experiments. Experiment 1 demonstrates how the GNN forecasting model can determine the optimal time to implement prescribed fires and focuses on the short-term pollution effect of prescribed fires. Experiment 2, on the other hand, focuses on quantifying the pollution impact of prescribed fires across months. In the rest of the section, we discuss each experiment in more detail:

\begin{figure}[!h]
  \centering
   \includegraphics[width=0.70\textwidth]{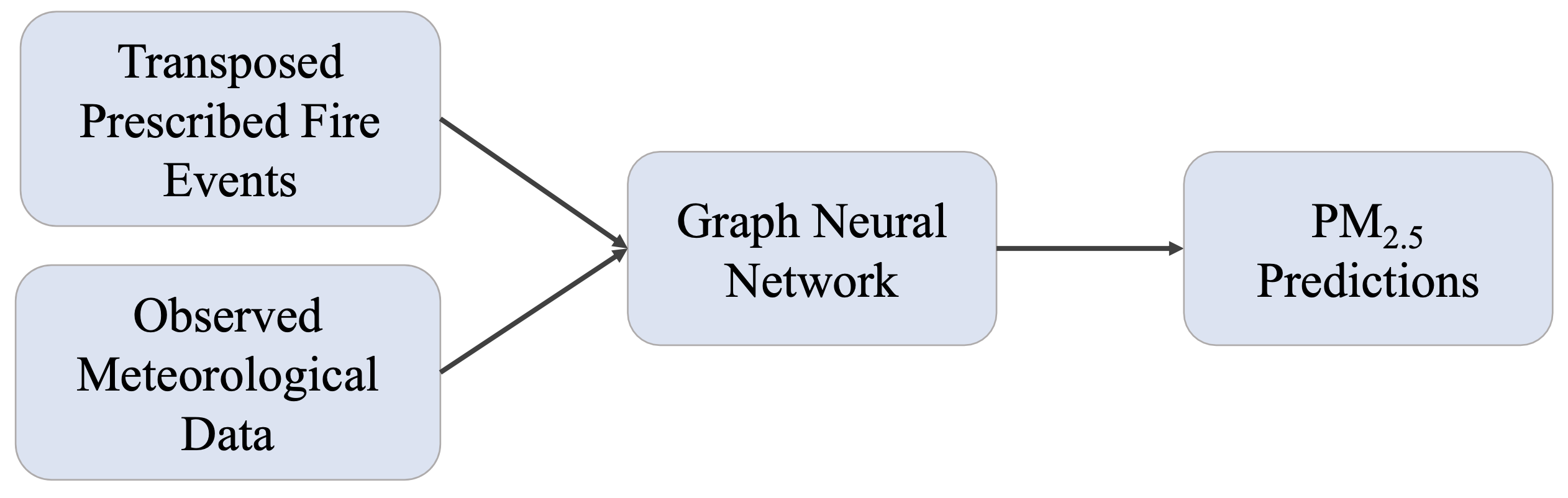}
  \caption{Prescribed fire simulation framework}
  \label{fig:pipeline}
\end{figure}

\subsection{Experiment 1: Minimizing Prescribed Fire PM$_{2.5}$ Impact} \label{exp1method}

\subsubsection{Methods}
To determine the optimal time to implement prescribed fires, we consider the immediate effect of prescribed fires. That is, we transpose the FRP values from actual prescribed fire events to target time points and add them to the observed FRP values at those points. As these FRP values are combined, they are aggregated using inverse distance and wind-based weighting, as outlined in Section \ref{dataset}.

\begin{figure*}
  \centering
   \includegraphics[width=\textwidth]{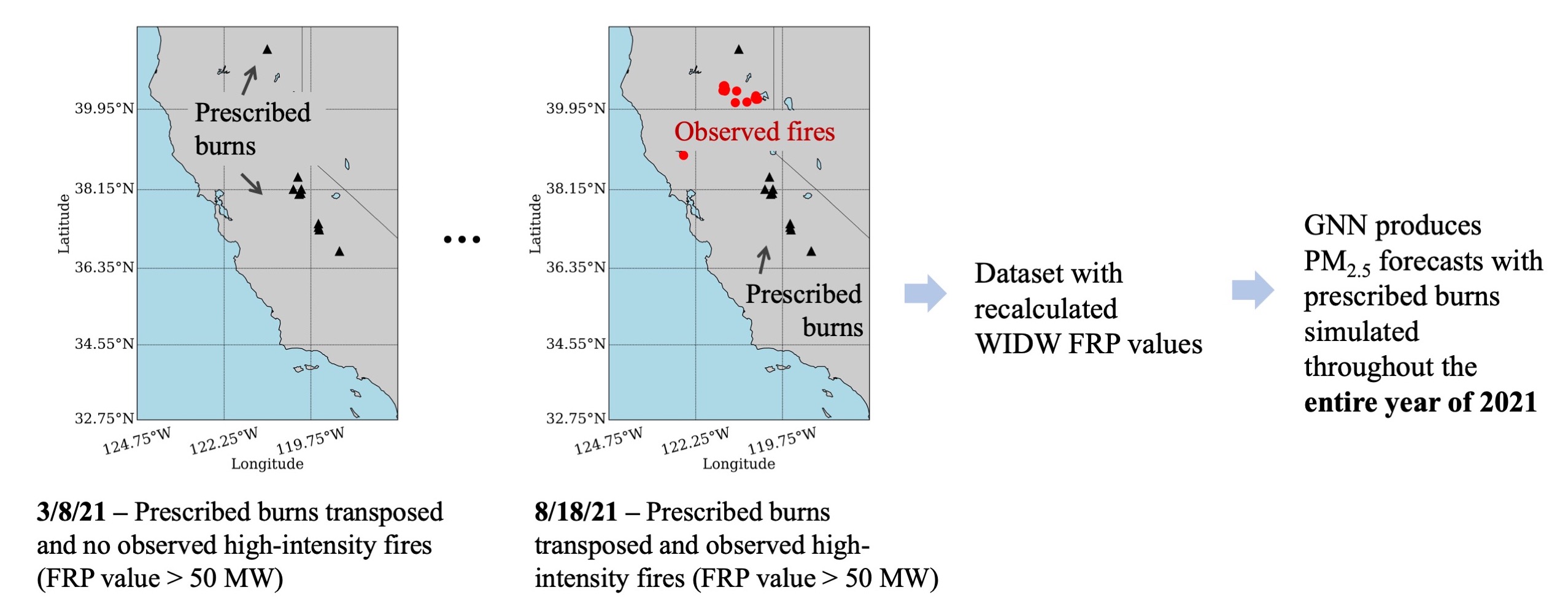}
    \caption{Schematic illustration of the methodology used in Experiment 1 to generate PM$_{2.5}$ predictions based on simulated prescribed burns and observed fire events throughout 2021}
    \label{fig:exp1method}
\end{figure*}

In this experiment, we transpose a window of time containing prescribed fires (1/3/21 - 1/15/21) to target times throughout the year 2021 at 24-hour time steps to simulate the air quality impacts of controlled burns, as shown in Figure \ref{fig:exp1method}. This window contains ten prescribed fires with burned areas above 100 acres. 

\subsubsection{Results}
As shown in Table \ref{table:exp1resulttable}, the month of August was the least optimal time to implement prescribed fires since it resulted in the most significant PM$_{2.5}$ concentration. As August is during the peak wildfire season, implementing prescribed fires would only exacerbate the already hazardous air quality. August’s mean PM$_{2.5}$ was 29.61\% greater than the average mean of other months, and August’s maximum was 44.27\% greater than the average maximum of other months. On the other hand, March, which had the lowest mean value, was found to be the most optimal month to implement prescribed fires. The mean and maximum values were calculated by averaging the mean and maximum PM$_{2.5}$ predictions of the locations whose PM$_{2.5}$ observations were $\geq$ 50 $\mu$g/${\rm m}^3$ during the window 1/3/21 - 1/15/21. As the PM$_{2.5}$ observations at those locations were elevated during 1/3/21 - 1/15/21, they were likely influenced by the fire events transposed across the year 2021.

\begin{table}
  \caption{Comparing the results of PM$_{2.5}$ predictions based on simulated prescribed fires in Experiment 1 (see text for more details) for each month of 2021}
  \centering
  \resizebox{\columnwidth}{!}{
  \begin{tabular}{l llllllllllll}
    \toprule
    & Jan     & Feb     & Mar     & Apr     & May     & Jun     & Jul     & Aug     & Sep     & Oct     & Nov     & Dec     \\
    \cmidrule(r){2-13}
    Mean ($\mu$g/${\rm m}^3$) & 15.62 & 15.64 & 14.69 & 15.18 & 14.76 & 15.92 & 18.44 & 21.73 & 18.55 & 16.95 & 19.94 & 18.73\\
    Max ($\mu$g/${\rm m}^3$) & 36.49 & 39.06 & 39.10 & 38.76 & 40.85 & 42.04 & 47.25 & 60.13 & 43.44 & 45.61 & 40.54 & 45.32 \\
    \bottomrule
  \end{tabular}
}
\label{table:exp1resulttable}
\end{table}

\subsection{Experiment 2: Quantifying Prescribed Fire PM$_{2.5}$ Trade-Off} \label{exp2method}

\subsubsection{Methods}
\begin{figure*}
  \centering
   \includegraphics[width=\textwidth]{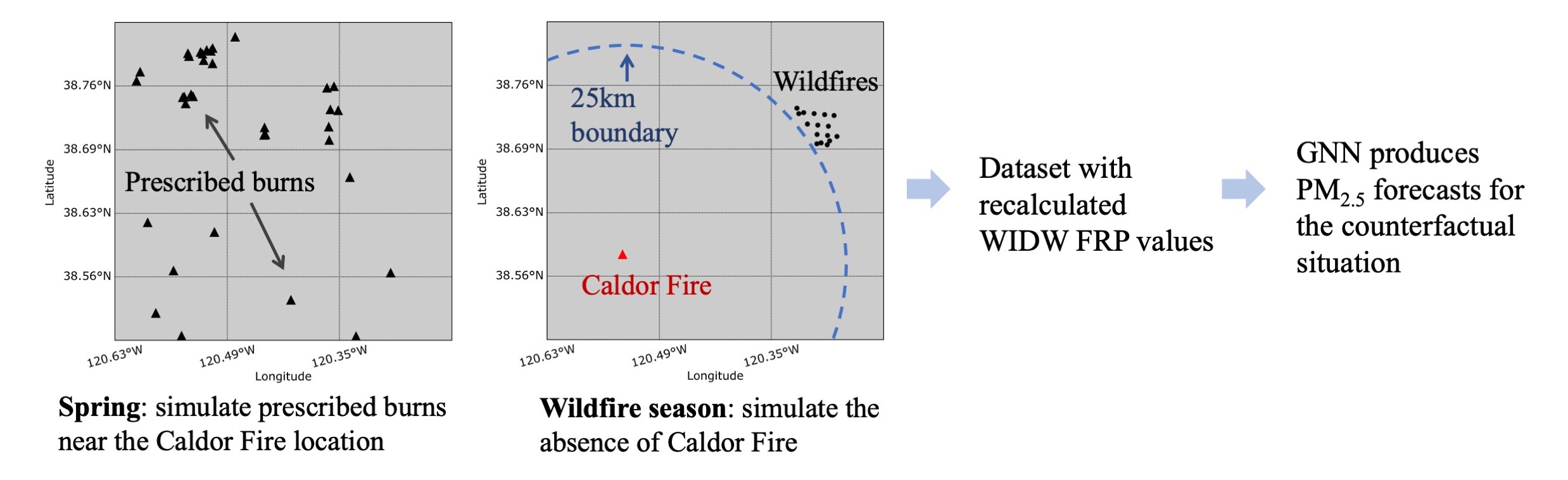}
    \caption{Schematic illustration of the methodology used in Experiment 2 for simulating prescribed burns during spring and the absence of the Caldor Fire during the wildfire season}
    \label{fig:exp2method}
\end{figure*}

This experiment aims to quantify the pollution trade-off of implementing prescribed fires by simulating the effect of controlled burns in 2021 at the location of the Caldor Fire, one of the largest Californian wildfires in 2021, as shown in Figure \ref{fig:exp2method}. We employ two simulation techniques: one corresponding to the immediate air quality impact of prescribed fires, and the other to the longer-term effect of prescribed burning, related to mitigating the emissions from a larger wildfire. 

For the first case, we simulated the effect of three historical prescribed fires, which were all within 20km of the 2021 Caldor Fire, were active from 3/21 - 5/31 in 2018, 2019, and 2020 respectively, and burned around 6,300 acres each. Since the Caldor Fire burned around 221,835 acres \cite{caldorcit}, we assume that preventing a fire of that scale would require a larger controlled burn. Thus, when creating the fire-related input variables, the FRP values from the prescribed fires are artificially increased by a factor of 100 and transposed together to 2021, thereby simulating large prescribed fires from 3/21/21 - 5/31/21. As described in Section \ref{exp1method}, the prescribed fires are transposed by combining the FRP values of the prescribed fires with the observed FRP values at the target time point, and then by aggregating those values using inverse distance weighting and the wind information.

In the latter case, we simulate the effect of controlled burns later in the year by excluding all FRP values within 25km of the Caldor Fire between 8/14/21 and 10/21/21, implicitly assuming that a prescribed fire implemented earlier in the year (or even during the previous one to two fire seasons) could effectively mitigate a large fire in the same location a few months later. To further remove the Caldor Fire influence, PM$_{2.5}$ values from 2018 are used as inputs instead of the Caldor-influenced, observed PM$_{2.5}$ values from 2021. 2018 PM$_{2.5}$ data is chosen because, in comparison to the other years, the 2018 fire season most closely resembles the 2021 fire activity without having fires at the Caldor Fire location.

The PM$_{2.5}$ predictions from this experiment’s counterfactual scenario are compared to baseline predictions derived using observed meteorological and fire inputs from 2021 without any prescribed fires around the Caldor Fire locations. 

\subsubsection{Results}
\begin{figure*}
  \centering
   \includegraphics[width=\textwidth]{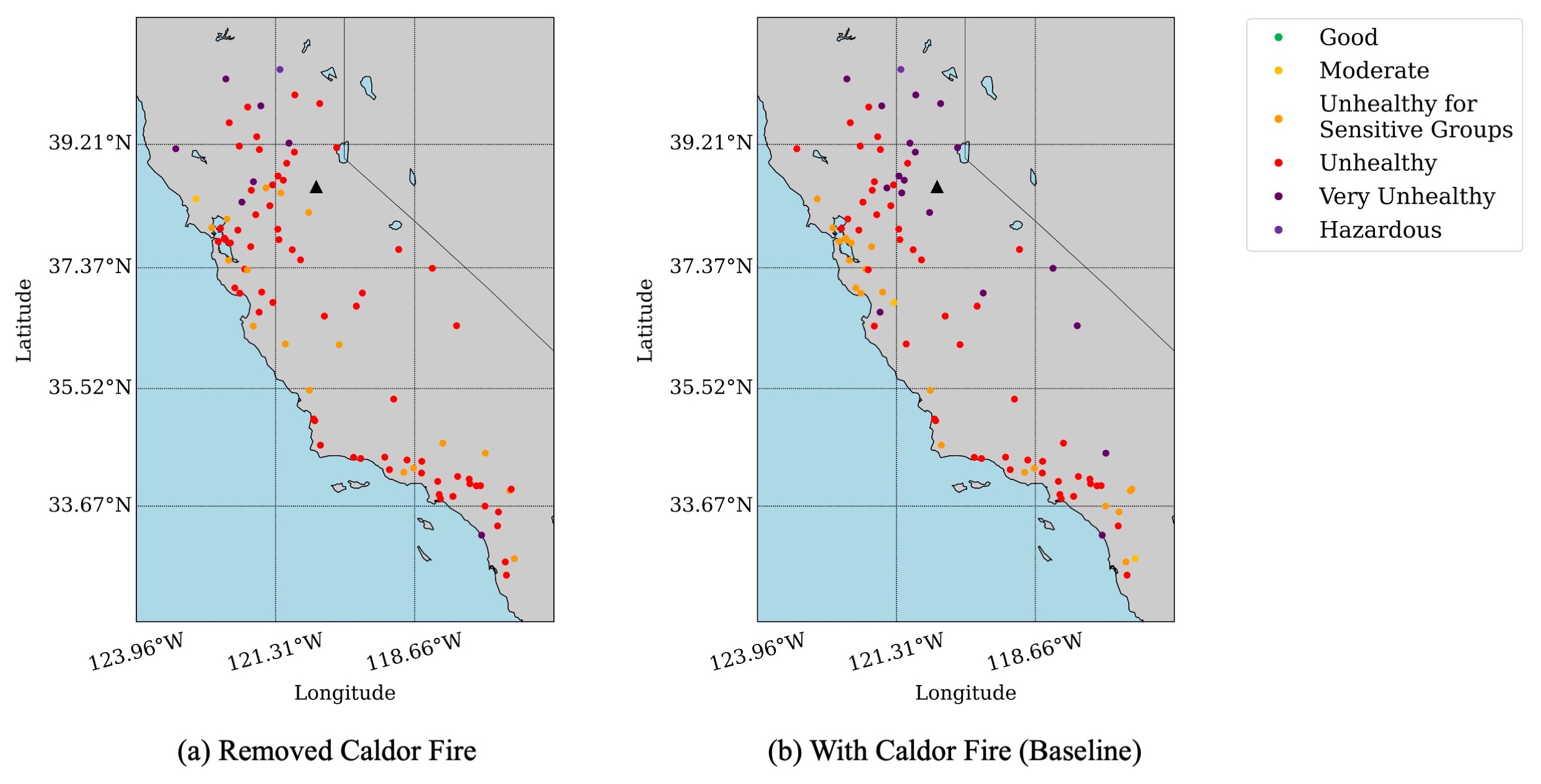}
    \caption{Maximum PM$_{2.5}$ predictions per site from 8/14/21 - 12/31/21 under condition (a) with prescribed burns at the Caldor Fire location during the spring and without Caldor Fire during the wildfire season and (b) without prescribed burns at the Caldor Fire location and with the Caldor Fire during the wildfire season}
    \label{fig:exp2caldor}
\end{figure*}

The results support that, though prescribed fires slightly increase PM$_{2.5}$ in the short term, the prescribed fires reduce future PM$_{2.5}$ resulting from wildfires. As shown in Table \ref{table:exp2resulttable}, the simulated prescribed burns led the mean of the PM$_{2.5}$ predictions to be increased by an average of 0.31 $\mu$g/${\rm m}^3$ and the maximum PM$_{2.5}$ prediction to be increased by 3.07\%. The mean and maximum values were calculated by averaging the mean and maximum PM$_{2.5}$ predictions of the 13 PM$_{2.5}$ monitor locations within 100km of the Caldor Fire. Table \ref{table:exp2resulttable} also quantifies that the maximum of the predictions with the Caldor Fire's influence removed was 52.85\% lower than the maximum of the baseline predictions. Thus, the magnitude of the immediate PM$_{2.5}$ increase from the prescribed fire was significantly lower than the magnitude of the PM$_{2.5}$ decrease experienced during the fire season. Furthermore, excluding the influence of the Caldor Fire reduced the number of days with an unhealthy daily average PM$_{2.5}$ concentration from a mean of 3.54 days to 0.70 days. The reduction in PM$_{2.5}$ pollution after excluding the Caldor Fire influence is illustrated in Figure \ref{fig:exp2caldor}, where the PM$_{2.5}$ monitoring sites are color-coded depending on the PM$_{2.5}$ pollution’s US AQI level.
 
\begin{table}
  \caption{Comparing the predicted PM$_{2.5}$ effect of simulated prescribed burns in Experiment 2 (see text for more details) with baseline PM$_{2.5}$ predictions}
  \label{table:exp2resulttable}
  \centering
  \fontsize{9}{11}\selectfont %
  \begin{tabular}{l ll ll}
    \toprule
    & \multicolumn{2}{c}{3/21/21 - 5/31/21}  & \multicolumn{2}{c}{8/14/21 - 10/21/21} \\
    \cmidrule(r){2-3} \cmidrule(r){4-5} 
    & \parbox{2.5cm}{\centering Simulated Prescribed Burn}     & \parbox{2.5cm}{\centering Without Prescribed Burn (Baseline)}     & \parbox{2.5cm}{\centering  Removed Caldor Fire}     & \parbox{2.5cm}{\centering With Caldor Fire (Baseline)} \\
    \cmidrule(r){2-3} \cmidrule(r){4-5} 
    Mean ($\mu$g/${\rm m}^3$) & 6.83 & 6.52 & 10.49 & 16.24 \\
    Max ($\mu$g/${\rm m}^3$) & 55.78 & 54.12 & 83.61 & 177.32 \\
    \bottomrule
  \end{tabular}
\end{table}

\section{Conclusion and Future Work}

This work produced hourly PM$_{2.5}$ predictions for California using a GNN model and demonstrated that the GNN outperformed other machine learning models like an MLP or LSTM, likely due to its spatio-temporal modeling capabilities. The temporally fine-grained, hourly PM$_{2.5}$ predictions can help people better plan their outdoor activities to stay healthy. Additionally, our work focused on exploring two novel applications of the GNN model: 1) producing fire-specific PM$_{2.5}$ forecasts, and 2) predicting the PM$_{2.5}$ for simulated fire events. 

To the best of our knowledge, our paper is the first to apply GNNs to the task of distinguishing the PM$_{2.5}$ pollutant concentration emitted from fires versus ambient sources. This is significant as machine learning has higher computational efficiency than chemical transport models (CTMs), and the GNN model has been shown to outperform other machine learning models for pollution forecasting.

Furthermore, to our understanding, this is also the first research paper to apply machine learning for simulating the PM$_{2.5}$ impact of prescribed fires, which is significant given the limitations of CTMs, as outlined above. A major contribution of this work is the prescribed fire simulation framework, which integrates prescribed fire simulations with GNN-based PM$_{2.5}$ predictions. Future work will focus on improving the fire simulation by incorporating physics-based modeling in the GNN framework. Our framework provides land managers and the fire service with a useful tool to minimize the PM$_{2.5}$ exposure of vulnerable populations, while also informing local communities of potential air quality impacts and beneficial trade-offs when implementing controlled burns.

\begin{Backmatter}


\paragraph{Funding Statement}
We acknowledge funding from NSF through the Learning the Earth with Artificial Intelligence and Physics (LEAP) Science and Technology Center (STC) (Award \#2019625). Jatan Buch, Kara Lamb, and Pierre Gentine were also supported in part by the Zegar Family Foundation.

\paragraph{Competing Interests}
None

\paragraph{Data Availability Statement}
Replication data and code can be found on GitHub: \url{https://github.com/kyleenliao/Prescribed_Fire_PM2.5_Simulation}

\paragraph{Ethical Standards}
The research meets all ethical guidelines, including adherence to the legal requirements of the study country.

\paragraph{Author Contributions}
Conceptualization: Kyleen Liao; Jatan Buch; Kara Lamb; Pierre Gentine. Methodology: Kyleen Liao; Jatan Buch; Kara Lamb; Pierre Gentine. Data curation: Kyleen Liao. Data visualization: Kyleen Liao. Writing original draft: Kyleen Liao. All authors approved the final submitted draft.


\end{Backmatter}


\newpage
\begin{thebibliography}{}
\bibitem{williams2019}
\textbf{Williams, A. P., Abatzoglou, J. T., Gershunov, A., Guzman‐Morales, J., Bishop, D. A., Balch, J. K., \& Lettenmaier, D. P.} (2019). Observed impacts of anthropogenic climate change on wildfire in California. \emph{Earth’s Future}, 7(8), 892–910. https://doi.org/10.1029/2019ef001210 

\bibitem{burke2021}
\textbf{Burke, M., Driscoll, A., Heft-Neal, S., Xue, J., Burney, J., \& Wara, M.} (2021). The changing risk and burden of wildfire in the United States. \emph{Proceedings of the National Academy of Sciences, 118}(2). https://doi.org/10.1073/pnas.2011048118 

\bibitem{who}
\textbf{World Health Organization.} (2022). \emph{Ambient (Outdoor) Air Pollution.} https://www.who.int/news-room/fact-sheets/detail/ambient-(outdoor)-air-quality-and-health

\bibitem{burke2023}
\textbf{Burke, M., Childs, M.L., de la Cuesta, B. et al.} The contribution of wildfire to PM$_{2.5}$ trends in the USA. \emph{Nature} (2023). https://doi.org/10.1038/s41586-023-06522-6

\bibitem{aguilera2021}
\textbf{Aguilera, R., Corringham, T., Gershunov, A. et al.} Wildfire smoke impacts respiratory health more than fine particles from other sources: observational evidence from Southern California. \emph{Nat Commun} 12, 1493 (2021). https://doi.org/10.1038/s41467-021-21708-0

\bibitem{wang2020PM25GNN}
\textbf{Wang, S., Li, Y., Zhang, J., Meng, Q., Meng, L., \& Gao, F.} (2020). PM$_{2.5}$-GNN. \emph{Proceedings of the 28th International Conference on Advances in Geographic Information Systems.} https://doi.org/10.1145/3397536.3422208

\bibitem{aguilera2023}
\textbf{Rosana Aguilera, Nana Luo, Rupa Basu, Jun Wu, Rachel Clemesha, Alexander Gershunov, Tarik Benmarhnia,} A novel ensemble-based statistical approach to estimate daily wildfire-specific PM$_{2.5}$ in California (2006–2020), \emph{Environment International}, Volume 171, 2023, 107719, ISSN 0160-4120, https://doi.org/10.1016/j.envint.2022.107719.

\bibitem{kelp2023}
\textbf{Kelp, M. M., Carroll, M. C., Liu, T., Yantosca, R. M., Hockenberry, H. E., \& Mickley, L. J.} (2023). Prescribed Burns as a tool to mitigate future wildfire smoke exposure: Lessons for states and Rural Environmental Justice Communities. \emph{Earth’s Future, 11}(6). 

\bibitem{mccaffrey2006}
\textbf{McCaffrey, Sarah M. 2006.} Prescribed fire: What influences public approval. In: Dickinson, Matthew B., ed. 2006. Fire in eastern oak forests: delivering science to land managers, proceedings of a conference; 2005 November 15-17; Columbus, OH. Gen. Tech. Rep. NRS-P-1. Newtown Square, PA: U.S. Department of Agriculture, Forest Service, Northern Research Station: 192-198.

\bibitem{askariyeh2020}
\textbf{Askariyeh, M. H., Khreis, H. \& Vallamsundar, S.} Air pollution monitoring and modeling. In Traffic-Related Air Pollut (eds Khreis, H. et al.) 111–135 (Elsevier, 2020).

\bibitem{byun2006}
\textbf{Byun, D. \& Schere, K. L.} Review of the governing equations, computational algorithms and other components of the models-3 community multiscale air quality (CMAQ) modeling system. \emph{Appl. Mech. Rev} 59, 51–76 (2006).

\bibitem{zaini2022}
\textbf{Zaini, N., Ean, L. W., Ahmed, A. N., Abdul Malek, M., \& Chow, M. F.} (2022). PM$_{2.5}$ forecasting for an urban area based on Deep Learning and Decomposition Method. \emph{Scientific Reports, 12}(1). https://doi.org/10.1038/s41598-022-21769-1 

\bibitem{rybarczyk2018}
\textbf{Rybarczyk, Y. \& Zalakeviciute, R.} Machine learning approaches for outdoor air quality modelling: A systematic review. \emph{Appl. Sci.} https://doi.org/10.3390/app8122570 (2018).

\bibitem{li2023}
\textbf{Li, L., Wang, J., Franklin, M. et al.} Improving air quality assessment using physics-inspired deep graph learning. \emph{npj Clim Atmos Sci} 6, 152 (2023). https://doi.org/10.1038/s41612-023-00475-3

\bibitem{carb}
\textbf{California Air Resources Board.} Air Quality and Meteorological Information System [internet database] available via https://www.arb.ca.gov/aqmis2/aqmis2.php. Accessed June 7, 2023.

\bibitem{epa}
\textbf{US Environmental Protection Agency.} Air Quality System Data Mart [internet database] available via https://www.epa.gov/outdoor-air-quality-data. Accessed June 7, 2023.

\bibitem{hersbach2020}
\textbf{Hersbach, H, Bell, B, Berrisford, P, et al.} The ERA5 global reanalysis. \emph{Q J R Meteorol Soc.} 2020; 146: 1999–2049. https://doi.org/10.1002/qj.3803

\bibitem{schroeder2014}
\textbf{Schroeder, W., Oliva, P., Giglio, L., \& Csiszar, I. A.} (2014). The new VIIRS 375 M active fire detection data product: Algorithm description and initial assessment. \emph{Remote Sensing of Environment}, 143, 85–96. https://doi.org/10.1016/j.rse.2013.12.008 

\bibitem{stekhoven2011}
\textbf{Stekhoven, D. J., \& Bühlmann, P.} (2011). Missforest—non-parametric missing value imputation for mixed-type data. \emph{Bioinformatics, 28}(1), 112–118. https://doi.org/10.1093/bioinformatics/btr597

\bibitem{calfire}
\textbf{Cal Fire.} Prescribed Burns [internet database] available via https://data.ca.gov/dataset/prescribed-burns. Accessed June 20, 2023.

\bibitem{reid2021}
\textbf{Reid, C.E., Considine, E.M., Maestas, M.M. et al.} Daily PM$_{2.5}$ concentration estimates by county, ZIP code, and census tract in 11 western states 2008–2018. \emph{Sci Data} 8, 112 (2021). https://doi.org/10.1038/s41597-021-00891-1

\bibitem{who2006}
\textbf{World Health Organization.} Regional Office for Europe \& Joint WHO/Convention Task Force on the Health Aspects of Air Pollution. (2006). Health risks of particulate matter from long-range transboundary air pollution. Copenhagen: WHO Regional Office for Europe. https://apps.who.int/iris/handle/10665/107691

\bibitem{mcclure2018}
\textbf{McClure, C. D., \& Jaffe, D. A.} (2018). US particulate matter air quality improves except in wildfire-prone areas. 
\emph{Proceedings of the National Academy of Sciences, 115}(31), 7901–7906. https://doi.org/10.1073/pnas.1804353115

\bibitem{caldorcit}
\textbf{Cal Fire.} Caldor Fire. Cal FIRE. https://www.fire.ca.gov/incidents/2021/8/14/caldor-fire/. Accessed October 18, 2023. 


\end{thebibliography}
\end{document}